\documentclass[12pt]{article}
\usepackage{amsfonts,amsmath,amssymb}
\usepackage{scalerel}[2016/12/29]
\usepackage{graphicx}
\usepackage{tikz}
\usepackage{float}
\usepackage{url}

\newcommand{\bb}{\hskip -0.1cm}
\newcommand{\bib}{\hskip -0.2mm}

\def\be{\begin{equation}}
\def\ee{\end{equation}}
\def\bea{\begin{eqnarray}}
\def\eea{\end{eqnarray}}

\usepackage{amsmath}
\usepackage{epsfig,graphicx}
\usepackage{graphicx}
\usepackage{tikz}
\def\be{\begin{equation}}
\def\ee{\end{equation}}
\def\bea{\begin{eqnarray}}
\def\eea{\end{eqnarray}}

\begin{document}

\thispagestyle{plain}

\title{\bf\Large }
\title{\Huge{Exclusion statistics for particles with a discrete spectrum}}

\author{St\'ephane Ouvry$^*$~ {\scaleobj{0.9}{\rm and}} \ Alexios P. Polychronakos$^\dagger$}

\date{\today}

\maketitle

\begin{abstract}
{We formulate and study the microscopic statistical mechanics of systems of particles with exclusion statistics {in} a  discrete one-body spectrum.
The statistical mechanics   of these systems can be expressed in terms of effective single-level grand partition
functions obeying a generalization of the  standard thermodynamic exclusion statistics equation of state. 
%Interestingly, two distinct such effective
%partition functions can be defined (``forward'' and ``backward'') obeying different equations of state but
%leading to identical statistical mechanics.
We derive explicit expressions for the thermodynamic potential in terms of microscopic cluster coefficients and
show that the mean occupation numbers of levels satisfy a nesting relation involving a number of
adjacent levels determined by the exclusion parameter. We apply the formalism to the harmonic
Calogero model and point out a relation with the  Ramanujan continued fraction identity and appropriate
generalizations.}
\end{abstract}

\noindent
* LPTMS, CNRS,  Universit\'e Paris-Sud, Universit\'e Paris-Saclay,\\ \indent 91405 Orsay Cedex, France; {\it stephane.ouvry@u-psud.fr}

\noindent
$\dagger$  Department of Physics, City College of New York and the Graduate Center of CUNY, New York, NY 10038, USA;
{\it apolychronakos@ccny.cuny.edu}
\vskip 1cm

\vfill
\eject

%\tableofcontents

\section{Introduction}

Anyons statistics \cite{LM}, and the related topics of fractional and exclusion statistics \cite{Haldane}, are enjoying renewed popularity since the 2020 announcement of experimental confirmations \cite{exp} of lowest Landau level (LLL)  anyon excitations with statistics $1/3$ in fractional quantum Hall
samples at filling $1/3$. {The statistics relevant to these results are abelian,} which is indeed the simplest and,
in principle, easiest to observe  among non conventional statistics, leaving aside possible more elaborate  nonabelian extensions. In this context, it is of interest to study the manifestation of exclusion statistics in its microscopic setting, and this is the aim of the present work.

The concept of exclusions statistics, as introduced by Haldane \cite{Haldane}, essentially holds only at the thermodynamic limit, or in situations where the Hilbert space consists of degenerate states as in the LLL of an external magnetic field. {  Its {\it ab initio} microscopic formulation (i.e., starting from  a 1-body spectrum and {filling it with particles obeying exclusion statistics},
 as can be done in the standard Bose and Fermi cases)  is, in general, impossible
for non integer (fractional) exclusion parameter $g$, since exact many-body states cannot be defined.} Microscopic concepts
become, in principle, accessible when the exclusion parameter $g$ is constrained to be an integer, but are again to a large extent ambiguous as they are not invariant under Hilbert space state reparametrizations.

This  is to be contrasted to  anyon statistics,   defined in 2 dimensions in terms of a  microscopic
$N$-anyon quantum Hamiltonian \cite{Wil}, with a statistical (exchange) anyon parameter $\alpha$ taking continuous values in $[0,2)$. Physics is periodic in $\alpha$ with period 2: $\alpha=0$ corresponds to bosons, $\alpha=1$ to fermions,{  and $\alpha=2$  again to bosons}. 
  It is  well known  that Haldane/exclusion statistics {and anyon
statistics} are intimately related when one considers anyons projected onto the LLL of an external magnetic field,   the 
so-called LLL-anyon model \cite{Ouv}.  Here exclusion statistics manifests itself for a  system of particles with a continuous degenerate $1$-body density of states in the LLL.

{Our present aim is  to focus on exclusion statistics in its most general microscopic setting}, i.e.,  for particles with  a
{discrete}  $1$-body  spectrum  $\epsilon_1, \epsilon_2, \ldots, \epsilon_q$. In this situation, exclusion statistics  requires a natural (and dynamically relevant) ordering of the $1$-body {energy eigenstates}, as, for example,
when a single quantum number {$k$} indexes them. This is typically the case for 1-dimensional systems,
but exclusion statistics can be relevant in more general settings provided that  a principal quantum number induces a natural
ordering of the  spectrum, as  it happens for example in the case of  the 2-dimensional LLL-anyon model properly regularized at long distances by a harmonic well.

{In the next section we will give} a summary of anyon statistics in the context
of the LLL-anyon model  \cite{Ouv} and will review the intimate relation of Haldane exclusion statistics to  LLL-anyon statistics. We will also consider the thermodynamics of the Calogero model \cite{NRBos,Isa} as yet another example of a microscopic realization of exclusion statistics. This will allow  for  a  general definition of the   
thermodynamics relations governing a gas of particles with exclusion statistics and a continuous $1$-body density of states.%,   for which the  exclusion parameter   can take any non negative real values}. 

{We will then move to the situation of present interest, the  statistical mechanics  of a gas of particles in
$q$ discrete 1-body energy levels  {$\epsilon_1 , \epsilon_2 , \ldots ,  \epsilon_q$ in a specific ordering} with an  integer exclusion parameter $g$. Bosons correspond to $g=0$, with no exclusion, while $g=1$ is the Fermi case where 
no more than one particle per quantum state is allowed. For $g$-exclusion, levels can again be occupied by at most one particle, but in addition at least $g\bib -\bib 1$ unoccupied levels must exist between any two occupied states.}
 
{Our main results, presented in section {\bf\ref{3}}, are that the statistical mechanics of these systems can be written as a generalization of the
thermodynamics relations {governing} LLL-anyon or Calogero particles, in terms of effective single-state grand partition
functions obeying a generalization of the standard exclusion statistics equation of state. Interestingly, two distinct such effective
partition functions can be defined, termed ``forward'' and ``backward'', obeying different equations of state but
leading to identical statistical mechanics.
We also define appropriate thermodynamic potentials and give their explicit expressions in terms of cluster coefficients that
involve sums over generalized partitions of the particle number. The mean occupation number of each 1-body level is
then expressed in terms of the effective grand partition functions and shown to satisfy nesting relations that involve
$g$ nearby levels. Finally, in section {\bf\ref{4}} the formalism is applied to the specific case of the harmonic Calogero system, and a relation to
the Ramanujan continued fraction identity is pointed out for $g=2$ and related generalizations for $g>2$.
We conclude with some directions for future research.}

\section{LLL-anyons and exclusion thermodynamics \label{2}}

We start with {a review of LLL-anyons in an isotropic harmonic trap of frequency $\omega$ and their intimate relation
with exclusions statistics. The spectrum of the $N$-anyon system is} 
\be
%E_N= ({\omega_t-\omega_c})\Bigl[\sum_{i=1}^N l_i +\alpha{N(N-1)\over 2}+N\Bigr]  +N\omega_c ~,\quad  0\le l_1\le l_2\le ...\le l_N
\label{eigen}
E_N= ({\omega_t-\omega_c})\Bigl[\sum_{i=1}^N l_i +\alpha{N(N-1)\over 2}\Bigr]  +N\omega_t ~,\quad  0\le l_1\le l_2\le ...\le l_N
\ee
where $\omega_c$ is half the cyclotron frequency, $\omega_t=\sqrt{\omega_c^2+\omega^2}$, the $l_i$ are 1-body
LLL angular momentum quantum numbers in 2 dimensions, and $\alpha$ is understood to be in the interval $[0,2)$. We stress that 
the harmonic well is introduced as a long distance regulator to split the degeneracy of the LLL (when $\omega=0$ the  LLL-anyon spectrum (\ref{eigen}) reduces trivially to $E_N=N\omega_c$) and {will be} taken to vanish in the thermodynamic limit. When $\alpha : 0\to 1$  the spectrum (\ref{eigen}) interpolates continuously
between the harmonic LLL-Bose and LLL-Fermi spectra. Going beyond $\alpha=1$, 
%one notes that if the anyon model is periodic in $\alpha$ of period $2$,  
 we note that due to the presence of the  magnetic field the Bose limit $\alpha\to 2^-$ differs from the standard Bose case $\alpha=0$  because of  some missing states. At $\alpha=2$  these missing states are restored thanks to   excited  states merging into the LLL ground state. 
 
 \iffalse Strictly speaking this implies  that the LLL projection is not  justified anymore at $\alpha=2$ resulting in a discontinuity at this point.]}{\color{green} [I still find this off the main point. There are two distinct questions: a) Why is (1) for $\alpha=2$ not the correct set of bosonic LLL states? (Answer: because they become a set of excited LLL states.). b) Where are the missing states? (Answer: they are formerly off-LLL states that join the LLL at $\alpha=2$). I see no sign that ``projecting to LLL is not justified any more''; in fact, I do not even know what this means. Chew on this until we talk tomorrow!]}\fi

{ The LLL-anyon spectrum (\ref{eigen}) is, in fact, identical in form to the spectrum of the 1-dimensional harmonic Calogero model with
interaction strength $\alpha ( \alpha -1)$. The
relation between the two models is well established (see \cite{polybis} for an explicit mapping) and the exclusion
statistics interpretation of their statistical mechanics is a common feature.
We will come back later in this section to  the well-known connection of the Calogero model to exclusion statistics in the thermodynamic limit $\beta\omega\to 0 $, and in section {\bf\ref{4}} we will examine in more
detail its microscopic statistical mechanics in the dicrete case  ($\beta\omega\ne 0 $) using the results of section {\bf\ref{3}}.} 

From the harmonic LLL-anyon  spectrum (\ref{eigen}) the $N$-body partition function  $Z_N $, grand partition function
${\cal Z}$ and cluster coefficients $b_n$, defined as
\be
{\cal Z}= \sum_{N=0}^{\infty} Z_N\, z^N ~,~~~\ln{\cal Z}=\sum_{n=1}^{\infty} b_n \,z^n
\nonumber\ee 
where $z$ is the fugacity, {can be calculated.
To probe the effect of the statistics, the cluster coefficients  $b_n$ were studied in \cite{Ouv} and found to leading order
in $\beta\omega$ (that is, for $\beta(\omega_t-\omega_c)$ small)  to be}
\be\label{cluster}
b_n= {1\over \beta(\omega_t-\omega_c)}{e^{-n\beta\omega_c}}\,{1\over n^2}\prod_{k=1}^{n-1}{k-n\alpha\over k} 
\ee
Taking then the thermodynamic limit $\beta\omega\to 0 $, {(which in the present case amounts to
$1/(n\beta^2\omega^2)\to V/\lambda^2$, where $V$ is the macroscopically large 2-dimensional volume --here, area-- of the system and $\lambda$ the
thermal wavelength), the LLL-anyon {thermodynamic potential} follows
\be
\ln {\cal Z}=N_L\ln y\label{corebisbis}
\ee
where $N_L=BV/\Phi_0$ is the LLL degeneracy, i.e., {the number of magnetic flux quanta in the volume  -- here, area -- of the system. The function $y$ was found to satisfy
\be
y- ze^{-\beta\omega_c}y^{1-\alpha}=1\;,
\label{corebisbisbis}\ee so that  
\be\nonumber
y=\sum_{N=0}^{\infty}{z^N e^{-\beta N\omega_c}}\prod_{k=2}^{N}{k-N\alpha\over k}
%y=1+ze^{-\beta\omega_c}+\sum_{N=2}^{\infty}{z^Ne^{-\beta N\omega_c}}\prod_{k=2}^{N}{k-N\alpha\over k}
\ee
(by definition $\prod_{k=k_1}^{k_2} (\dots\bib ) = 1$ when $k_2<k_1$).
 
Using $  {\cal {Z}}=y^{N_L}$, { it was deduced \cite{Ouv}}
\[
Z_N={e^{-\beta{ N\omega_c}}}{N_L\prod_{k=2}^{N}{k+N_L-N\alpha-1\over k}}
\]
$Z_N$ is  the LLL-anyon $N$-body partition function for $N$ degenerate anyons at energy $N\omega_c$, {thus
identifying their degeneracy as}
\be
{ N_L\prod_{k=2}^{N}{k+N_L-N\alpha-1\over k} = ~}
{N_L\over N!}{(N+N_L-N\alpha-1)!\over (N_L-N\alpha)!}\label{count1}
\ee
{(factorials for fractional argument are defined in terms of the corresponding $\Gamma$-functions.)}

Let us now allow $\alpha$ to take  integer values  beyond the interval $[0,2)$:  the degeneracy (\ref{count1}) counts   
the  number of ways to put $N$ particles in  $N_L$ degenerate  quantum states in a circular configuration such that there
are at least 
$\alpha-1$ empty  states {between any two} occupied states \cite{Poly}. {This is the hallmark of exclusion statistics  with exclusion parameter $\alpha$.
{In particular, $\alpha= 2$ describes a Bose gas but with nontrivial  $\alpha=2$ exclusion.}

\noindent Indeed, Haldane exclusion statistics postulates that
given  $G$ single-particle states already populated by  $N-1$ particles 
the number  of quantum states available for an additional  ${N^{\text {th}}}$ particle is {$G-(N-1)g$}
{(this is heuristic, and somewhat misleading \cite{Poly}) where $g$ is  the  exclusion parameter. $g$ would need to be an integer for this to be meaningful, but this
requirement can be dropped in the thermodynamic limit where $G$ and $N$ become large. Starting} from the standard Bose degeneracy  for $N$ bosons in $G$ quantum states 
\[
{(N+G-1)!\over N!(G-1)!}
\]
{Haldane encoded} exclusion by replacing  $G$ by $ G-(N-1)g$ {to propose} the $N$-body exclusion degeneracy
\be\label{count2}
{(N+G-(N-1)g-1)!\over N!(G-(N-1)g-1)!}
\ee
When $g$ is {a positive integer,} this is the number of ways to put on a line $N$ particles in $G$ quantum states  in  such a way  there are at least $g-1$ empty  states in between two occupied states. {This is the same as the
LLL-anyon counting (\ref{count1}) discussed  above upon setting $g=\alpha$ and $G= N_L$ in the Haldane counting
and placing the states on a line rather than a circle, which is} irrelevant in the thermodynamic limit  where  $G=N_L\to\infty$. So Haldane exclusion statistics is identical to LLL-anyon statistics provided that the anyonic exchange statistical parameter $\alpha$ is allowed to take integer values beyond $[0,2)$. Not surprisingly, {in view} of this intimate relation between LLL-anyon statistics  and Haldane exclusion statistics, the LLL-anyon thermodynamic (\ref{corebisbis}) and (\ref{corebisbisbis})  can be  directly recovered  \cite{wu}   from the Haldane Hilbert space counting  (\ref{count2}).

In conclusion, exclusion/LLL-anyon thermodynamics amounts to
\be \label{corebis}
\ln {\cal Z}=N_L\ln y\;\;\;{,}\;\;\;
y- ze^{-\beta\omega_c}y^{1-\alpha}=1
\ee
Let us  focus on the  mean particle number $\bar N = z {\partial \ln Z/\partial z}$ or, equivalently, on the  LLL filling factor \be \nu={\bar N\over N_L}=z{\partial \ln y\over \partial z}
\nonumber\ee
Using $y- ze^{-\beta\omega_c} y^{1-\alpha}=1$ {we can obtain}
\[ y= 1+{\nu\over 1-\alpha\;\nu}
\]
and therefore also
\[
ze^{-\beta\omega_c}={\nu\over (1+
(1-\alpha)\nu)^{1-\alpha}(1-\alpha\;\nu)^{\alpha}}
\]
and from
$ \ln {\cal Z}=N_L\ln y$ we obtain the equation of state
\[\ln {\cal Z}=\beta P V=N_L\ln(1+{\nu\over 1-\alpha\;\nu})\]  
with a critical filling  $\nu={1/ \alpha}$  where the pressure diverges \cite{Ouv}. Interestingly, at the critical filling
the  $N$-body LLL-anyon wave  function  is nondegenerate 
\be
\psi=\prod_{i<j}(z_i-z_i)^{\alpha}e^{-\omega_c \sum_{i=1}^N\bar z_i z_i/2}
\label{laugh}\ee
which coincides with {the} Laughlin wavefunction  when  $\alpha=2m+1$, $m=1,2,\ldots$, encoding  $\alpha$-exclusion  for an incompressible $N$-anyon state in the LLL.

{The above thermodynamics readily generalizes to exclusion statistics systems with arbitrary 1-body  density of
states $\rho(\epsilon)$. In the thermodynamic limit we can apply relations (\ref{corebis}) for states around energy
$\epsilon$, with the number of states $N_L$ substituted by $\rho(\epsilon) d\epsilon$, obtaining
\be 
\ln{\cal Z}=\int_0^{\infty}\rho(\epsilon)\ln {y}(\epsilon)\;d\epsilon\quad{,}\quad {y}(\epsilon)- ze^{-\beta\epsilon}{y}(\epsilon)^{1-\alpha}=1
\label{core}\ee
The LLL-anyon result (\ref{corebis}) is recovered for $\rho(\epsilon) = N_L\delta(\epsilon-\omega_c)$. Similarly,
for the mean occupation number $n(\epsilon)$ per level at energy $\epsilon$ and the mean particle number $\bar N$
\[
n(\epsilon)=z{\partial \ln {y}(\epsilon)\over \partial z} ~~,~~~
\bar N=\int_0^{\infty}\rho(\epsilon) n(\epsilon)\;d\epsilon
\]
we obtain, in view of (\ref{core}),
\bea
{y}(\epsilon)&=&1+ {n(\epsilon)\over 1-\alpha\; n(\epsilon)} \nonumber\\
ze^{-\beta\epsilon}&=&{n(\epsilon)\over [1+(1-\alpha)n(\epsilon)]^{1-\alpha}[1-\alpha\; n(\epsilon)]^{\alpha}}
\label{n}\eea
and  therefore 
\be 
\ln {\cal Z}=\int_0^{\infty}\rho(\epsilon)\ln\bigg(1+ {n(\epsilon)\over 1-\alpha\; n(\epsilon)}\bigg) \;d\epsilon
\label{ZZ}\ee
}

 A question naturally arises about the  existence of other microscopic quantum models with the same kind of statistics. One known example, as already mentioned, is the Calogero model.  %It is indeed  tempting to look at  (\ref{corebis}) as  relevant not only to the LLL-anyon $1$-body density of states 
The harmonic Calogero model on the 1-dimensional line  with inverse-square 2-body interactions and a confining
harmonic potential is described by the Hamiltonian
\be\label{Ham}
 H_N= -\frac{1}{2}\sum_{i=}^N \frac{\partial^2}{\partial x_i^2} - 
\alpha(1-\alpha)\sum_{i<j}\frac{1}{(x_i-x_j)^2}
+\frac{1}{2}\,\omega^2\, \sum_{i=1}^N x_i^2
\ee
Its spectrum is given by
\be
E_N= \omega\big[\sum_{i=1}^N l_i +\alpha{N(N-1)\over 2}+{ N\over 2}  \big]\quad 0\le l_1\le l_2\le ...\le l_N
\label{eigenbis}\ee
where the $l_i$ are now 1-dimensional ``pseudo-excitation numbers'' labeling the  $1$-body harmonic eigenstates.
{This spectrum directly follows from the the LLL-anyon spectrum (\ref{eigen}) upon letting $\omega_c \to 0$, i.e. in the absence of the external magnetic field, up to a trivial $N$-body energy shift $N\omega/2$}.
Note that $\omega$ now plays the role of a 1-dimensional long-distance regulator in the Calogero case,
with the difference that the limit $\omega \to 0$ does
not lead anymore to an infinitely degenerate LLL but, rather, to free 1-dimensional particles with {generalized} statistics manifesting through
their scattering phase shift \cite{NRBos}.

All the thermodynamic considerations presented in this section in the context of LLL anyons apply equally to Calogero
particles. In particular, when $\omega_c=0$ the cluster coefficients (\ref{cluster}) in the  thermodynamic limit $\beta\omega\to 0$ lead  to (\ref{core}, \ref{n}, \ref{ZZ}) with density of states
\[
\rho(\epsilon)={L\over\pi\sqrt{2\epsilon}}
\]
 i.e., a free 1-dimensional density of states on a space of macroscopically large length $L$.

{ So the $2$-dimensional LLL-anyon and the $1$-dimensional Calogero models both share the same  exclusion statistics/thermodynamics (\ref{core}, \ref{n}, \ref{ZZ}).
 This  is of course not surprising: as already stressed above one can show \cite{polybis} that the $1$-dimensional Calogero model is a particular projection of the $2$-dimensional anyon model with  same exclusion statistics and    a free 1-body density of states  on the line stemming in the LLL from the dimensional reduction
 $\lim_{\omega_c\to 0} N_L\delta(\epsilon-\omega_c)={L\over\pi\sqrt{2\epsilon}}$ induced by the vanishing magnetic field.}
 
All said and done,  (\ref{core}, \ref{n}, \ref{ZZ})  can be viewed in full generality as the defining thermodynamic relations for particles with exclusion statistics $\alpha$ and a continuous 1-body density of states $\rho(\epsilon)$. However, specific dynamical
systems manifesting these thermodynamics, where  exclusion statistics is {\it microscopically} realized in terms of $N$-body quantum  Hamiltonians, are limited to the two cases   above -- the LLL-anyon and the Calogero models. In the
next section we consider in general such systems, defined through a set of 1-body energy levels and
an exclusion rule in filling them for many-body states, and derive their exact {\it microscopic statistical mechanics},
rather than their thermodynamics. 

\section{Exclusion statistics for a discrete 1-body spectrum \label{3}}
\vskip -0.8cm
We now turn to exclusion statistics for a discrete density of states, that is for a {1-body spectrum $\epsilon(k)$, $k=1, \ldots,q$. As stressed in the Introduction, we assume a natural ordering of levels 
$\epsilon(1),\epsilon(2),\ldots,\epsilon(q)$ in
terms of the principal quantum number $k$ in $\epsilon(k)$, which is relevant to the definition of exclusion}. The 
Boltzmann factor for the energy level $\epsilon(k)$ is
\be\nonumber 
s(k)=e^{-\beta\epsilon(k)}
\ee
{We call $s(k)$} the spectral function. Our focus is to derive relations analogous to (\ref{core}, \ref{n}, \ref{ZZ}) for particles with exclusion statistics in the discrete spectrum above.  
\subsection{$g=2$ \label{g=2}}
As a warmup, we start with the simplest case beyond Fermi statistics, i.e., $g=2$ exclusion.

The  $N$-body partition function is, by definition,
\bea
Z_N 
 = \sum_{k_1=1}^{q-2N+2} \sum_{k_2=1}^{k_1} \cdots \sum_{k_{N}=1}^{k_{N-1}}
s(k_1+2N-2)
s(k_2+2N-4) \cdots s(k_{N-1}+2)
s(k_{N})\nonumber\label{Z}
\eea
where the cumulative $+2$ shifts in the arguments of the spectral function  enforce $g=2$ exclusion: adjacent 1-body levels $k$ and $k+1$ cannot be populated.
The grand partition function follows as
\[{\cal{Z}}_{1,q}=1+\sum_{N=1}^{(q+1)/2}Z_Nz^N\]
{(the indices $1,q$ refer to the first and last levels in the spectrum). The cluster expansion of the grand potential is
\be
\ln {\cal{Z}}_{1,q}=\sum_{n=1}^{\infty}b_nz^n
%=-\sum_{n=1}^{\infty}(-z)^n   \hskip -0.3 cm\sum_{{l_1, l_2, \ldots, l_{j}\atop { \rm composition}\;{\rm of}\;n}\atop j\le q} \hskip -0.4cm 
%c_2(l_1,l_2,\ldots,l_{j} )\sum _{k=1}^{q-j+1} s^{l_{j}}(k+j-1)\cdots s^{l_2}(k+1) {s}^{l_1}(k)
\nonumber
\ee}
with the cluster coefficients $b_n$ calculated to be
\be b_n=(-1)^{n-1}   \hskip -0.3 cm\sum_{{l_1, l_2, \ldots, l_{j}\atop { \rm composition}\;{\rm of}\;n}\atop j\le q} \hskip -0.4cm 
c_2(l_1,l_2,\ldots,l_{j} )\sum _{k=1}^{q-j+1} s^{l_{j}}(k+j-1)\cdots s^{l_2}(k+1) {s}^{l_1}(k)\label{helpp}
\ee
In (\ref{helpp}),  the sum is  over all compositions (i.e., ordered partitions) of the integer $n$, with the number of parts $j$ of a given composition, by definition smaller than or equal to $n$, also constrained to be smaller than or equal to $q$, the number of available $1$-body quantum states. 
The  combinatorial coefficients $c_2(l_1,l_2,\ldots,l_{j})$ are  \cite{latt} 
\be
c_2(l_1,l_2,\ldots,l_{j}) = \frac{{l_1+l_2\choose l_1}}{l_1+l_2}\;\; l_2\frac{{l_2+l_3\choose l_2}}{l_2+l_3}\;\cdots \;\; l_{{j}-1}\frac{{l_{{j}-1}+l_{j}\choose l_{{j}-1}}}{l_{{j}-1}+l_{j}}\;.\nonumber
\ee
{By rearranging the sums in (\ref{helpp}), the cluster coefficient can also take the alternative form  
 \be\nonumber b_n =(-1)^{n-1}\sum_{k=1}^{q}\sum_{{l_1, l_2, \ldots, l_{j}\atop { \rm composition}\;{\rm of}\;n}\atop j\le q-k+1} \hskip -0.4cm 
c_2(l_1,l_2,\ldots,l_j ) s(k+j-1)^{l_j}\cdots s(k+1)^{l_2} s(k)^{l_1}\;.\nonumber\ee }
\vskip -0.3cm
A useful observation \cite{emeis} is that ${\cal{Z}}_{1,q}$ can be expressed
as the secular determinant of the off-diagonal $(q+1)\times (q+1)$ matrix
\be 
H_{1,q} = \begin{pmatrix}
0 & -1 & 0 & \cdots & 0 & 0\\s(1) & 0 &-1 & \cdots & 0 & 0\\
0 & s(2) & 0 & \cdots & 0 & 0\\
\vdots & \vdots & \vdots & \ddots & \vdots & \vdots\\
0 & 0 & 0 & \cdots & 0 & -1 \\
0 & 0 & 0 & \cdots & s(q) & 0 \\
\end{pmatrix}\label{matrix}\ee
%where, on each column below the entry $1$ located just above the diagonal, there is  a vanishing entry, a manifestation of $g=2$ exclusion in this matrix representation
(this $H_{1,q}$ is one of several equivalent choices).
Specifically,
\be\nonumber 
{\cal{Z}}_{1,q}=\det(1_{q+1}+z^{1/2}H_{1,q})
\label{det1q}\ee
%\iffalse{\red (Here is an alternative that fully displays the matrix and makes its expansion easy to see:)\\
%A useful observation \cite{emeis} is that ${\cal{Z}}_{1,q}$ can be expressed
%as the determinant of the secular off-diagonal $(q+1)\times (q+1)$ matrix
%\be 
%H_{1,q}(z) = \begin{pmatrix}
%1 & -z & 0 & \cdots & 0 & 0\\s(1) & 1 &-z & \cdots & 0 & 0\\
%0 & s(2) & 1 & \cdots & 0 & 0\\
%\vdots & \vdots & \vdots & \ddots & \vdots & \vdots\\
%0 & 0 & 0 & \cdots & 1 & -z \\
%0 & 0 & 0 & \cdots & s(q) & 1 \\
%\end{pmatrix}\label{matrix}\ee
%(this $H_{1,q}$ is one of several equivalent choices).
%That is,
%\be
%{\cal{Z}}_{1,q}=\det[H_{1,q} (z)]
%\label{det1q}\ee
%(I know you will hate it, but wrote it all the same!)}\fi
{We also define the general truncated grand partition functions ${\cal{Z}}_{k,k'}$ ($1\le k\le k'\le q$) as
\be {\cal{Z}}_{k,k'}=\det(1 +z^{1/2}H_{k,k'})
\label{recur}\ee
with $H_{k,k'}$ the truncated $(k'-k+2)\times (k'-k+2)$ matrix 
\be 
H_{k,k'} = \begin{pmatrix}
0 & -1 & 0 & \cdots & 0 & 0\\s(k) & 0 &-1 & \cdots & 0 & 0\\
0 & s(k+1) & 0 & \cdots & 0 & 0\\
\vdots & \vdots & \vdots & \ddots & \vdots & \vdots\\
0 & 0 & 0 & \cdots & 0 & -1 \\
0 & 0 & 0 & \cdots & s(k') & 0 \\
\end{pmatrix}\nonumber\ee
i.e.,  the matrix (\ref{matrix}) for a truncated 1-body spectrum $\epsilon(k), \epsilon(k+1), \ldots, \epsilon(k')$ starting at level $k$ and ending at level $k'$. We also define that, trivially,  ${\cal{Z}}_{k,k'}=1$  when $k>k'$.

The  grand partition  function ${\cal{Z}}_{k,k'}$ can be expressed as
\[{\cal{Z}}_{k,k'}=1+\sum_{N=1}^{(k'-k+2)/2}{Z_{N}}(k,k') z^N
\]
where ${Z_{N}}(k,k')$  stands  for the  $g=2$ exclusion  $N$-body partition function  for the truncated 
1-body spectrum  $\epsilon(k), \epsilon(k+1), \ldots, \epsilon(k')$, i.e.,
 \[Z_{N}(k,k')= \sum_{k_1=k}^{k'-2N+2} \sum_{k_2=k}^{k_1} \cdots \sum_{k_{N}=k}^{k_{N-1}}
s(k_1+2N-2)
s(k_2+2N-4) \cdots s(k_{N-1}+2)
s(k_{N})\;. \] It also follows that
\be
\ln {\cal{Z}}_{k,k'}=-\sum_{n=1}^{\infty}(-z)^n\hskip -0.4cm \sum_{{l_1, l_2, \ldots, l_{j}\atop { \rm composition}\;{\rm of}\;n}\atop j\le  k'-k+1 } \hskip -0.4cm 
c_2(l_1,l_2,\ldots,l_{j} )\sum _{l=k}^{k'-j+1} s^{l_{j}}(l+j-1)\cdots s^{l_2}(l+1) {s}^{l_1}(l)
\label{cluskk}\ee
where the number of parts $j$ is now bounded by  the number $k' -k+1$ of available quantum states in the truncated spectrum (here, as well as in all other similar cluster expressions which will appear below).}

Expanding the determinant (\ref{det1q}) in terms of its first row we obtain
\be  
{\cal{Z}}_{1,q}={\cal{Z}}_{2,q}+z s(1){\cal{Z}}_{3,q}\label{help}\ee
The recursion (\ref{help})  is  self explanatory and could have been written {directly}: because of $g=2$  exclusion, the full grand partition function is the sum of a grand partition for a $1$-body spectrum starting at level $\epsilon(2)$ {with level $\epsilon(1)$
empty, and of $zs(1)$ (level $\epsilon (1)$ filled) times the one} starting at level $\epsilon(3)$. This is the basis for a recursion scheme: {expanding the determinants  ${\cal Z}_{k,q}$ and ${\cal Z}_{1,k}$, as defined in (\ref{recur}), in terms of
their first and last row, respectively, yields the recursion relations
\be 
{\cal{Z}}_{k,q}={\cal{Z}}_{k+1,q}+z s(k){\cal{Z}}_{k+2,q} ~,~~~{\cal{Z}}_{1,k}={\cal{Z}}_{1,k-1}+z s(k){\cal{Z}}_{1,k-2}
\label{recurk}\ee
We now introduce ``forward" and ``backward'' effective single-level grand partition functions ${y_+}(k)$ and ${y_-}(k)$, respectively, as
\be 
y_+ (k)={{\cal{Z}}_{k,q}\over {\cal{Z}}_{k+1,q}}  ~;~~~{y_-}(k)={{\cal{Z}}_{1,k}\over {\cal{Z}}_{1,k-1}}
\label{etbien}\ee
In terms of them, the full grand partition function achieves a product form 
\be
 {\cal{Z}}_{1,q}=\prod_{k=1}^q {y_+}(k)=\prod_{k=1}^q {y_-}(k)
 \label{product}\nonumber\ee
Moreover, the recursion relations (\ref{recurk}) imply the nesting relations
\be {y_\pm}(k)- {z s(k) \over {y_\pm}(k\pm 1)}=1% ~,~~~{y_-}(k)- {z s(k) \over {y_-}(k-1)}=1 
\label{nest} \ee
where ${y_+}(0)={y_-}(0)={y_+}(q+1)={y_-}(q+1)=1$ is understood.
We also note that, as a consequence of (\ref{cluskk}, \ref{etbien}), the cluster expansions of ${y_+}(k),{y_-}(k)$ are
\bea
\ln {y_+}(k)&=&-\sum_{n=1}^{\infty}(-z)^n  \bb\bb\bb\sum_{{l_1, l_2, \ldots, l_{j}\atop { \rm composition}\;{\rm of}\;n}\atop j\le q-k+1} \hskip -0.4cm 
c_2(l_1,l_2,\ldots,l_{j} ) s(k)^{l_1} s(k+1)^{l_2} \cdots  s(k+j-1)^{l_j}~~ \nonumber\\
\ln {y_-}(k)&=&-\sum_{n=1}^{\infty}(-z)^n \bb\bb\bb\sum_{{l_1, l_2, \ldots, l_{j}\atop { \rm composition}\;{\rm of}\;n}\atop j\le k} \hskip -0.4cm 
c_2(l_1,l_2,\ldots,l_{j} ) s(k-j+1)^{l_j} \cdots s(k-1)^{l_2} s(k)^{l_1} ~~\label{good}\nonumber
\eea}

We reach the conclusion that for a discrete 1-body spectrum %$\epsilon(1),\epsilon(2),\ldots,\epsilon(q)$
the statistical mechanics of  $g=2$ exclusion particles {amounts to
\bea
\ln {\cal{Z}}_{1,q}&=&\sum_{k=1}^q \ln {y_\pm}(k) ~,~~~  {y_\pm}(k)-{z s(k) \over {y_\pm}(k\pm 1)}= 1 \nonumber
%\text{or,}~~ \ln {\cal{Z}}_{1,q}&=&\sum_{k=1}^q \ln {y_-}(k) ~,~~~  {y_-}(k)-{z s(k) \over {y_-}(k-1)}= 1
\nonumber\eea
These are indeed (\ref{core}) when $\alpha=2$ but with the proviso that, because of the discreteness  of the  spectrum, $\ln {\cal{Z}}_{1,q}$ is now a discrete sum instead of a continuous integral, and a  discrete shift $k\to k\pm 1$ materializes in the argument of ${y_+}(k)$ or ${y_-}(k)$.}

Turning to the {mean} particle number {$\bar N$, it is given by}
\be\nonumber
\bar N = z{\partial \ln {\cal {Z}}_{1,q}\over \partial z}=\sum_{k=1}^q n_k\ee
where the mean occupation number  $n_k$ of the  energy  level $\epsilon(k)$  is, by definition,
%\iffalse{\red\be
%n_k = z{\partial \ln {y_+}(k)\over \partial z} 
%\ee}
%(by convention $n_k=0$  when $k<1$ and $k>q$)
%\fi
\be
n_k = s(k){\partial \ln{\cal Z}_{1,q}\over \partial s(k)}\label{nk}\nonumber
\ee
\vskip -0.5cm
\noindent Using the expansion relation
\be{\cal Z}_{1,q}=zs(k){\cal Z}_{1,k-2}{\cal Z}_{k+2,q}+{\cal Z}_{1,k-1}{\cal Z}_{k+1,q} \;,\nonumber \ee 
(another self-explanatory identity of which both (\ref{help}) and (\ref{recurk}) are special cases)
{$n_k$ can be expressed directly in terms of truncated grand partition functions}
\be
n_k = zs(k) \,{{\cal Z}_{1,k-2}{\cal Z}_{k+2,q}\over  {\cal Z}_{1,q}} ~~\Leftrightarrow~~ 1-n_k={{\cal Z}_{1,k-1}{\cal Z}_{k+1,q}\over  {\cal Z}_{1,q}}\;. \label{neni}\nonumber
\ee
{From this we can obtain the nesting relation for $n_k$ (see Appendix for the proof)}
\be
z s(k) = {n_k (1-n_k ) \over (1-n_{k-1}-n_k )(1-n_k -n_{k+1} )} 
\label{nss}\ee
{and using (\ref{etbien}) we can express ${y_+}(k)$ and ${y_-}(k)$ in terms of $n_k$}
\be \label{ykfakenk} 
{y_\pm}(k)=1+ {n_k\over 1- n_k - n_{k\mp 1}}% ~,~~~{y_-}(k)=1+ {n_k\over 1-n_k - n_{k+1}} 
\ee
%\iffalse
%{\red 
%(\ref{nest}) implies
%\be \label{ykfakenkbis} 
%{y_+}(k)=1+ {n_k\over 1- n_{k}-n_{k+1}}
%\ee
%which can be used to derive a nesting relation for $n_k$
%\be\label{zsfakenk}
%zs(k)={n_k\over 1-n_k-n_{k+1}} \cdot {1-n_{k+2}\over  1-n_{k+1}-n_{k+2}}
%\ee
%and  finally  the thermodynamical potential
%\fi
and finally the thermodynamic potential
%\iffalse 
% \be 
%\ln {\cal Z}_{1,q}=\sum_{k=1}^q\ln\bigg(1+ {n_k\over 1-n_k-n_{k+1}}\bigg) 
%\ee
%\fi
\be 
\ln {\cal Z}_{1,q}=\sum_{k=1}^q\ln\bigg(1+ {n_{k}\over 1-n_{k-1}-n_{k}}\bigg)=\sum_{k=1}^q\ln\bigg(1+ {n_{k}\over 1-n_{k}-n_{k+1}}\bigg) 
\label{life}\ee
(\ref{nss}, \ref{ykfakenk}, \ref{life}) are the {generalization} of the $\alpha=2$ LLL-anyon/Calogero models  (\ref{n}, \ref{ZZ}) thermodynamics relations  for the discrete spectrum at hand, i.e., with  the discrete shifts {$k\to k\pm 1$} in the arguments of $n$. (\ref{life})  is the equation of state  of a gas of particles with exclusion $g=2$ and populating  discrete energy levels $\epsilon(k)$  whose  occupation numbers  $n_k$  are constrained by {$n_{k}+n_{k+1}\le 1$}.

%We note that instead of the "forward" ${y_+}(k)$ one could as well has defined, starting from the bottom of the spectrum $\epsilon(1)$, the "backward"  single level grand partition function
%\be{\tilde{y}}(k)={{\cal {Z}}_{1,k}\over {\cal {Z}}_{1,k-1}}=1+ {n_k\over 1- n_{k}-n_{k+1}}
%\nonumber\ee where  ${\tilde{y}}(k)$   satisfies to the nesting relation
%\be {\tilde{y}}(k)- {z s(k) \over {\tilde{y}}(k-1)}=1\label{99} \ee with
%\be\ln {\cal Z}_{1,q}=\sum_{k=1}^q\ln{\tilde{y}}(k) 
%\nonumber\ee 

Note that the nesting relation  (\ref{nss}) does not allow for finding the $n_k$ in an iterative way, starting either at $k=1$ or at $k=q$,
since already for $k=1$ it involves $n_1$ {\it and} $n_2$ and similarly for $k=q$. On the other hand, (\ref{nest}) allows
the calculation of ${y_+}(k)$ and $ {y_-} (k)$ iteratively, starting from $k=1$ for ${y_-} (k)$ and $k=q$ for ${y_+}(k)$:
\bea
&&{y_-} (1)=1+z s(1) ~, ~~ {y_-} (2) = 1+{z s(2) \over 1+z s(1)}~ , \dots  \nonumber \cr
&&{y_+}(q) = 1+z s(q)~, ~~{y_+}(q\bib-\bib\bib 1) =   1+{z s(q\bb -\bb 1) \over 1+z s(q)}~ , \dots \cr
\nonumber \eea
From these {and (\ref{nss})} we can express $n_k$ as
\be
n_k = {{y_+}(k) -1 \over {y_+}(k) +  {y_-} (k-1) -1} = {{y_-} (k) -1 \over {y_-} (k) + {y_+}(k+1) -1}\nonumber
\ee

\subsection{General $g$}
For general integer exclusion parameter $g$ the $N$-body partition function reads
\bea
Z_N 
 = \sum_{k_1=1}^{q-gN+g} \sum_{k_2=1}^{k_1} \cdots \sum_{k_{N}=1}^{k_{N-1}}
s(k_1+gN-g)
s(k_2+gN-2g) \cdots s(k_{N-1}+g)
s(k_{N})\label{Zbis}
\eea
where the $+g$ shift in the nested sum indices enforces $g$-exclusion. The grand partition function  ${\cal{Z}}_{1,q}$ follows as
\[
{\cal{Z}}_{1,q}=1+\sum_{N=1}^{(q+g-1)/g}Z_Nz^N
\] 
and the cluster expansion  as
\be\ln {\cal{Z}}_{1,q}%=\sum_{n=1}^{\infty}b_n z^n
=-\sum_{n=1}^{\infty}(-z)^n \hskip -0.5cm \sum_{{l_1, l_2, \ldots, l_{j}\atop \text{g-composition}\;{\rm of}\;n}\atop j\le q} \hskip -0.5cm 
c_g(l_1,l_2,\ldots,l_{j} )\sum _{k=1}^{q-j+1} s^{l_{j}}(k+j-1)\cdots s^{l_2}(k+1) {s}^{l_1}(k)~~\label{clusg}\ee
which is the $g$-generalization of (\ref{helpp}).
{Here the sum is over all $g$-compositions \cite{latt} of the integer $n$   which are defined as the usual compositions but where now up to {$g-2$ successive $l_i$ can be zero}. The  $c_g(l_1,l_2,\ldots,l_{j} )$ are    given  as
\bea
{c_g (l_1,l_2,\ldots,l_{j})} &=& {{{(l_1+\dots +l_{g-1}-1)!\over l_1! \cdots l_{g-1}!}~
\prod_{i=1}^{j-g+1} {l_i+\dots +l_{i+g-1}-1 \choose l_{i+g-1}}}}\nonumber\\
&=& {{\prod_{i=1}^{j-g+1} (l_i + \dots + l_{i+g-1} -1)! \over \prod_{i=1}^{j-g} (l_{i+1} + \dots +l_{i+g-1} -1 )! } }\prod_{i=1}^j {1\over l_i!} \nonumber
\eea}

As {before we can express} ${\cal{Z}}_{1,q}$
as the  secular determinant of a $(q+g-1)\times (q+g-1)$ matrix
%\iffalse
%\be 
%H_{1,q} = \begin{pmatrix}
%0 & 1 & 0 & \cdots & 0 & 0\\
%\vdots & 0 &1 & \cdots & 0 & 0\\
%0 & \vdots & 0 & \cdots & 0 & 0\\
%(-1)^{g-1}s(1) & 0 & \vdots & \ddots & \vdots & \vdots\\
%\vdots & (-1)^{g-1}s(2) & 0 & \cdots & 0 & 1 \\
%\vdots & 0 & (-1)^{g-1}s(3) & \cdots & (-1)^{g-1}s(q) & 0 \\
%0 & 0 & (-1)^{g-1}s(3) & \cdots & (-1)^{g-1}s(q) & 0 \\
%\end{pmatrix}\ee\fi
\be \label{matrixbis}
H_{1,q} = \left(
\begin{array}{ccccccccc}
0 & -1 & 0 & \cdots & 0 & 0 & \quad 0 \quad & \cdots & \quad 0 \quad \\
0 & 0 & -1 & \cdots & 0 & 0 & 0 & \cdots & 0 \\
\vdots & \vdots & \vdots & \ddots & \vdots & \vdots & \vdots & \ddots & \vdots \\
s(1) & 0 & 0 & \ddots & -1 & 0 & 0 & \cdots & 0 \\
0 & s(2) & 0 & \ddots & 0 & -1 & 0 & \cdots & 0 \\
0 & 0 & s(3) & \ddots & 0 & 0 & -1 & \cdots & 0 \\
\vdots & \vdots & \vdots & \ddots & \vdots & \vdots & \vdots & \ddots & \vdots \\
0 & 0 & 0 & \cdots & s(q-1) & 0 & 0 & \cdots &- 1 \\
0 & 0 & 0 & \cdots & 0 & s(q) & 0 & \cdots & 0 \\
\end{array} \right)
\ee
%\iffalse
%here represented for $q=8$ and $g=4$
%\be \nonumber
%\left(
%\begin{array}{ccccccccccc}
%0 & 1 & 0 & 0 & 0 & 0 & 0 & 0  \;\;& 0  \;\;& \;\;\;\;0 & \;\;\;\;0 \\
%0 & 0 & 1 & 0 & 0 & 0 & 0 & 0 \;\; & 0  \;\;&\; \;\;\;0 &\;\; \;\;0 \\
%0 & 0 & 0 & 1 & 0 & 0 & 0 & 0  \;\;& 0  \;\;& \;\;\;\;0 &\; \;\;\;0 \\
%-s(1) & 0 & 0 & 0 & 1 & 0 & 0 & 0  \;\;& 0 \;\; &\; \;\;\;0 &\;\;\;\; 0 \\
%0 & -s(2) & 0 & 0 & 0 & 1 & 0 & 0  \;\;& 0  \;\;&\; \;\;\;0 &\;\; \;\;0 \\
%0 & 0 & -s(3) & 0 & 0 & 0 & 1 & 0  \;\;& 0  \;\;&\;\;\;\;0 &\; \;\;\;0 \\
%0 & 0 & 0 & -s(4) & 0 & 0 & 0 & 1  \;\;& 0  \;\;&\;\; \;\;0 &\;\; \;\;0 \\
%0 & 0 & 0 & 0 & -s(5) & 0 & 0 & 0  \;\;& 1  \;\;&\;\;\; \;0 &\; \;\;\;0 \\
%0 & 0 & 0 & 0 & 0 & -s(6) & 0 & 0  \;\;& 0  \;\;&\; \;\;\;1 &\;\;\; \;0 \\
%0 & 0 & 0 & 0 & 0 & 0 & -s(7) & 0  \;\;& 0  \;\;&\; \;\;\;0 &\; \;\;\;1 \\
%0 & 0 & 0 & 0 & 0 & 0 & 0 & -s(8) \;\; & 0  \;\;&\; \;\;\;0 & \;\;\;\;0\\
%\end{array}\;\;\; \right)
%\ee
%\fi
where there are now $g-1$ successive vanishing diagonals, realizing $g$-exclusion in the matrix representation.
Then
\be\nonumber 
{\cal{Z}}_{1,q}=\det(1+z^{1/g}H_{1,q})
\ee

Proceeding as in the $g=2$ case, {we define the general truncated grand partition functions ${\cal{Z}}_{k,k'}$
 ($1\le k\le k'\le q$) as
\be {\cal{Z}}_{k,k'}=\det(1 +z^{1/g}H_{k,k'})
\label{recurbis}\nonumber\ee
with $H_{k,k'}$ the $(k'-k+g)\times (k'-k+g)$ matrix (\ref{matrixbis}) for the truncated 1-body spectrum
$\epsilon(k), \epsilon(k+1), \ldots, \epsilon(k')$ starting at level $k$ and ending at level $k'$, and, as before, ${\cal{Z}}_{k,k'}=1$  when $k>k'$.
The  grand partition  function ${\cal{Z}}_{k,k'}$ can be expressed as
\[{\cal{Z}}_{k,k'}=1+\sum_{N=1}^{(k'-k+g)/g}{Z_{N}}(k,k') z^N
\]
where ${Z_{N}}(k,k')$  stands  for the truncated $g$-exclusion $N$-body partition function
\[{Z_{N}}(k,k') = \sum_{k_1=k}^{k'-g N+g} \sum_{k_2=k}^{k_1} \cdots \sum_{k_{N}=k}^{k_{N-1}}
s(k_1+gN-g)
s(k_2+gN-2g) \cdots s(k_{N-1}+g)
s(k_{N})\] 

We can again set up a recursion scheme by expanding the determinants  ${\cal Z}_{k,q}$ and ${\cal Z}_{1,k}$ in terms of
their first and last row, respectively, obtaining the recursion relations
\be 
{\cal{Z}}_{k,q}={\cal{Z}}_{k+1,q}+z s(k){\cal{Z}}_{k+g,q} ~,~~~
{\cal{Z}}_{1,k}={\cal{Z}}_{1,k-1}+z s(k){\cal{Z}}_{1,k-g}
\label{recurg}
\ee
of clear $g$-exclusion statistics origin. 
We define forward and backward effective single-level grand partition functions ${y_+}(k)$ and ${y_-}(k)$ as
in the $g=2$ case (\ref{etbien}), and the full grand partition function is again expressed as their product (\ref{product}).
The cluster expansion for ${\cal{Z}}_{k,k'}$ is as in (\ref{clusg}) but for the truncated spectrum, that is
\be
\ln {\cal{Z}}_{k,k'}=-\sum_{n=1}^{\infty}(-z)^n \hskip -0.4cm \sum_{{l_1, l_2, \ldots, l_{j}\atop \text{g-composition}\;{\rm of}\;n}\atop j\le k'-k+1} \hskip -0.4cm 
c_g(l_1,l_2,\ldots,l_{j} )\sum _{l=k}^{k' -j+1} s^{l_{j}}(l+j-1)\cdots s^{l_2}(l+1) {s}^{l_1}(l)
\label{cor}\ee
leading to expressions
analogous to (\ref{good}) for the logarithm  $\ln {y_+}(k)$ and $\ln{y_-}(k)$.
}

%Again, a recursion for ${\cal{Z}}_{k,q}$ follows as 
%\be  
%{\cal{Z}}_{k,q}={\cal{Z}}_{k+1,q}+z s(k){\cal{Z}}_{k+g,q}
%\label{recurgk}\ee
%\iffalse and
%\be\ln {\cal{Z}}_{k,q}=-\sum_{n=1}^{\infty}(-1)^n z^n\sum_{l_1, l_2, \ldots, l_{j}\atop { \rm g-composition}\;{\rm of}\;n} \hskip -0.4cm 
%c_g(l_1,l_2,\ldots,l_{j} )\sum _{k'=k}^{q-j+1} s^{l_{j}}_{k'+j-1}\cdots s^{l_2}_{k'+1} {s}^{l_1}_{k'}\nonumber\ee\fi
The recursion relation (\ref{recurg})  implies the nesting relations {
\be 
{y_\pm}(k)-{z s(k) \over \prod_{i=1}^{g-1}{y_+}(k\pm i)}=1
% ~,~~~
%{y_-}(k)-{z s(k) \over \prod_{i=1}^{g-1}{y_-}(k-i)}=1 
\label{atreat} \ee}
We reach the conclusion that the statistical mechanics for $g$-exclusion particles in a discrete 1-body spectrum $\epsilon(1),\epsilon(2),\ldots,\epsilon(q)$ amounts to
\bea
\ln {\cal{Z}}_{1,q}&=&\sum_{k=1}^q \ln {y_\pm}(k) ~,~~~  {y_\pm}(k)-{z s(k) \over \prod_{i=1}^{g-1}{y_\pm}(k\pm i)}= 1 \nonumber\\
%\text{or,}~~ \ln {\cal{Z}}_{1,q}&=&\sum_{k=1}^q \ln {y_-}(k) ~,~~~  {y_-}(k)-{z s(k) \over \prod_{i=1}^{g-1}{y_-}(k-i)}=1 
\label{discrexclus}\nonumber\eea
which is again basically (\ref{core})  with $\alpha$  replaced by $g$, but with a discrete sum instead of a continuous integral  and the discrete shifts $k\to k\pm1\ldots\to k\pm(g-1)$  in the argument of ${y_+}(k)$ and ${y_-}(k)$.

Likewise, {because of the  identity 
\be{\cal Z}_{1,q}={\cal Z}_{1,k-1}{\cal Z}_{k+g-1,q}+\sum_{i=k}^{k+g-2}zs(i){\cal Z}_{1,i-g}{\cal Z}_{i+g,q} \;,\nonumber \ee
of obvious exclusion statistics origin (which can also be obtained by expanding the secular determinant with respect
to its $k^{\text{th}}$ row),  the mean occupation number $n_k$ defined in (\ref{nk}) becomes}
\be
n_k = zs(k) {{\cal Z}_{1,k-g}{\cal Z}_{k+g,q}\over  {\cal Z}_{1,q}} \nonumber
\ee
and satisfies the nesting relation (see Appendix for the proof)
%\[
%z s(k)={n_k\over 1-\sum_{i=0}^{g-1} n_{k-i}}\prod_{j=1}^{g-1} { 1-\sum_{i=1}^{g-1} n_{k+j-i}\over 1-\sum_{i=0}^{g-1} n_{k+j-i}}
%\]
\[
z s(k)={n_k}\, {\prod_{j=1}^{g-1} \left(  1-\sum_{i=1}^{g-1} n_{k+j-i}\right) \over \prod_{j=0}^{g-1} 
\left( 1-\sum_{i=0}^{g-1} n_{k+j-i} \right)}
\]
{and combining with (\ref{atreat}) we obtain the expressions for ${y_+}(k)$ and ${y_-}(k)$ {
  \[  {y_\pm}(k)  = 1 +  {n_{k} \over 1-\sum_{j=0}^{g-1}n_{k\mp j}}% ~,~~~  {y_-}(k)  = 1 +  {n_{k} \over 1-\sum_{j=0}^{g-1}n_{k+j}}
\]
leading to the expression for the thermodynamic potential
\be \ln{ \cal Z}_{1,q}=\sum_{k=1}^q\ln\bigg(1+ {n_k\over 1-\sum_{i=0}^{g-1} n_{k-i}}\bigg) 
=\sum_{k=1}^q\ln\bigg(1+ {n_k\over 1-\sum_{i=0}^{g-1} n_{k+i}}\bigg) 
\nonumber\ee}
again in close similarity to (\ref{n}, \ref{ZZ})  for $\alpha$  replaced by $g$.  When the occupation numbers are such that for any given $k$ their sum over $g$ neighboring levels $\sum_{i=0}^{g-1} n_{k-i}=1$, 
{the system} has  reached the maximal critical filling {allowed} by exclusion statistics.

\section{The harmonic Calogero model \label{4}}
As an illustration of the formalism developed in the previous section, let us focus  on  the  $N$-body Calogero harmonic spectrum (\ref{eigenbis}), here again  for convenience shifted by $N\omega/2$, not any more in the thermodynamic limit  $\beta\omega\to 0$  as in Section ({\ref{2}), where the harmonic well played the role of a long-distance regulator, but keeping $\beta\omega$  physical and finite. This is an example of a exclusion-$g$ statistics system with a discrete 1-body spectrum, arising from a {\it microscopic} $N$-body Hamiltonian with spectral function
$s(k)=x^k$, $k=1,2,\ldots,\infty$,  with $x=e^{-\beta\omega}$, i.e., a harmonic linear $1$-body spectrum.
We could  focus as well on the $N$-body LLL-anyon harmonic spectrum (\ref{eigen})  with  spectral function  \be s(k)=x^k x_c\label{monLLL}\ee  where $x=e^{-\beta(\omega_t-\omega_c)}$  and $x_c=e^{-\beta\omega_c}$. 

\subsection{Number of states $q$ finite}

 {Assume, now}, that the number of 1-body states is finite, that is, $k=1,2,\ldots,q$. The integer $q$ is an effective high-energy cutoff which is not actually needed for the
consistency of the results; later on, the $q\to\infty$ limit will be taken. The $N$-body spectrum (\ref{eigenbis}), with $\alpha$ traded for $g$ and the $l_i$ bounded by $q\bib-\bib (N-1)g$,  yields a partition function which is nothing but $Z_N$  in  (\ref{Zbis})  for  the spectral function $s(k)=x^k$ discussed above. \iffalse It reads 
\be 
Z_N= x^{N + g N (N - 1)/2} \prod_{j=1}^N{1 - x^{j + q - 1 - (N - 1) g}\over 1 - x^j}
\nonumber\ee
for {$0 \le N \le (q +g- 1)/g$}. {\color{green} [This is basically a repetition -- decide if we keep it] \fi The grand partition function follows as
\be
{ \cal Z}_{1,q}=1+\sum_{N=1}^{(q+g - 1)/g }z^Nx^{N + g N (N - 1)/2} \prod_{j=1}^N{1 - x^{j + q - 1 - (N - 1) g}\over 1 - 
x^j}
\nonumber\ee}
In this specific situation, where $s(k)=x^k$, {we see that ${\cal Z}_{k,k'}$  is actually ${\cal Z}_{1,q}$ but with $q\to k'-k+1$ and $z\to z x^{k-1}$
\be { \cal Z}_{k,k'}=1+\sum_{N=1}^{(k' -k+g)/g }z^N x^{N + g N (N - 1)/2} \prod_{j=1}^N{x^{k-1} - x^{j + k' - 1 - (N - 1) g}\over 1 - x^j}
\label{coeff0}\nonumber\ee}
From the cluster expansion  (\ref{cor}) we obtain
\bea\ln { \cal Z}_{k,q}&=& -\sum_{n=1}^{\infty}\;(-z)^n \hskip -0.3cm \sum_{{{l_1, l_2, \ldots, l_{j}\atop \text{g-composition}\;{\rm of}\;n}\atop j\le q-k+1}} \hskip -0.3cm c_g(l_1,l_2,\ldots,l_{j} ) \;\sum_{l=k}^{q-j+1}x ^{n l} x^{\sum_{i=1}^j(i-1)l_i} \nonumber\\
%&=& {\red -\sum_{n=1}^{\infty}\;(-z)^n\; {x^{ n k}-x^{n(q-j+2)}\over 1-x^n}\hskip -0.3cm \sum_{{{l_1, l_2, \ldots, l_{j}\atop \text{g-composition}\;{\rm of}\;n}\atop j\le q-k+1}} \hskip -0.3cm c_g(l_1,l_2,\ldots,l_{j} ) \;x^{\sum_{i=1}^j(i-1)l_i}}\\
&=& {-\sum_{n=1}^{\infty} {(-z)^n \over 1-x^n}\hskip -0.3cm \sum_{{{l_1, l_2, \ldots, l_{j}\atop \text{g-composition}\;{\rm of}\;n}\atop j\le q-k+1}} \hskip -0.3cm  \left[x^{ n k}\bib -x^{n(q-j+2)}\right]  c_g(l_1,l_2,\ldots,l_{j} ) \;x^{\sum_{i=1}^j(i-1)l_i}}
\label{coeff} \eea
{and similarly
\be
\ln { \cal Z}_{1,k} =-\sum_{n=1}^{\infty} {(-z)^n \over 1-x^n}\hskip -0.3cm \sum_{{{l_1, l_2, \ldots, l_{j}\atop \text{g-composition}\;{\rm of}\;n}\atop j\le k}} \hskip -0.3cm  \left[x^n -x^{n(k-j+2)}\right]  c_g(l_1,l_2,\ldots,l_{j} ) \;x^{\sum_{i=1}^j(i-1)l_i}
\label{coefk}\nonumber
\ee
which imply the cluster expansion for the forward and backward effective single-level grand partition functions }
\bea
\ln {y_+}(k)&=&-\sum_{n=1}^{\infty}\;(-z x^{ k})^n\sum_{{l_1, l_2, \ldots, l_{j}\atop \text{g-composition}\;{\rm of}\;n}\atop j \le q-k+1}c_g(l_1,l_2,\ldots,l_{j} ) \; x^{\sum_{i=1}^j(i-1)l_i}  \label{coeffbis} \cr
\ln {y_-}(k)&=&-\sum_{n=1}^{\infty}\;(-z x^{ k})^n\sum_{{l_1, l_2, \ldots, l_{j}\atop \text{g-composition}\;{\rm of}\;n}\atop j \le k}c_g(l_1,l_2,\ldots,l_{j} ) \; x^{-\sum_{i=1}^j(i-1)l_i}  \label{coefftris} 
\eea
{The above expressions are remarkably similar, differing in the sign of the exponent and in the allowed $g$-compositions.
Note that the expression for $y_- (k)$ does not involve $q$, since it looks ``back'' towards $k=1$, and is, thus, universal.}

\subsection{Number of states $q\to
\infty$: the harmonic Calogero model}

{ Clearly in the $q\to\infty$ limit, i.e., the   harmonic Calogero model,  the constraint on the number of parts $j \le q-k+1$ in (\ref{coeff}) or (\ref{coeffbis}) disappears:   (\ref{coeff}) for $k=1$   is  the harmonic Calogero thermodynamic potential
\be 
\ln {\cal Z}_{1,q}= -\sum_{n=1}^{\infty}\;(-z)^n {x^n\over 1- x^n}\sum_{{l_1, l_2, \ldots, l_{j}\atop \text{g-composition}\;{\rm of}\;n}}c_g(l_1,l_2,\ldots,l_{j} ) \; x^{\sum_{i=1}^j(i-1)l_i}
\nonumber\ee
with the harmonic Calogero cluster coefficients
\be
b_n=(-1)^{n-1}{1\over 1- x^n}\sum_{{l_1, l_2, \ldots, l_{j}\atop \text{g-composition}\;{\rm of}\;n}}c_g(l_1,l_2,\ldots,l_{j} ) \; x^{\sum_{i=1}^j i \;l_i}\;.
\label{clusterbis}\ee
The $b_n$ in (\ref{clusterbis})  encode $g$-exclusion statistics arising from the microscopic $N$-body  harmonic Calogero model, i.e., from its Hamiltonian (\ref{Ham}) and $N$-body spectrum (\ref{eigenbis}).

Let us   consider the {small $\beta\omega$ limit} to recover the cluster coefficients of Section {\ref{2}}. {For small}
$\beta\omega$,  i.e., $x=e^{-\beta\omega}\simeq 1-\beta\omega$, the cluster coefficient $b_n$ in (\ref{clusterbis}) becomes
\be b_n= (-1)^{n-1} {1\over n \beta\omega}\;\sum_{{l_1, l_2, \ldots, l_{j}\atop \text{g-composition}\;{\rm of}\;n}} \;c_g(l_1,l_2,\ldots,l_{j} )\nonumber\ee  
Using \cite{polybis} \[\sum_{{l_1, l_2, \ldots, l_{j}\atop \text{g-composition}\;{\rm of}\;n}} \;c_g(l_1,l_2,\ldots,l_{j} )={{g n\choose n}\over gn}\]
it simplifies to
\be  b_n= (-1)^{n-1}{1\over \beta\omega}{{gn\choose n}\over gn^2}\nonumber\ee 
and  indeed coincides, when $g$ is traded {for} $\alpha$, with  (\ref{cluster}) when  $\omega_c=0$. In order to recover the full cluster coefficient (\ref{cluster}), i.e,  with $\omega_c\ne 0$, {we have to use the spectral function (\ref{monLLL}), that is, put $x= e^{-\beta(\omega_t-\omega_c)}$ in (\ref{clusterbis}) and multiply by $x_c^n$}.

{We remark} that in (\ref{clusterbis})
the coefficients of  the polynomial
\be\nonumber 
\sum_{{l_1, l_2, \ldots, l_{j}\atop \text{g-composition}\;{\rm of}\;n}} c_g(l_1,l_2,\ldots,l_{j} ) \; x^{\sum_{i=1}^j(i-1)l_i} \ee
are, when multiplied by  $n$, a $g$-generalization of the coefficients OEIS A227532 and the related A227543, in relation to
the Ramanujan continuous fraction.  Indeed, {denoting
\be 
H_g(z,x)=  { \cal Z}_{2,\infty}(-z/x,x) = 1+\sum_{N=1}^{\infty} z^N x^{gN(N-1)/2}\prod_{j=1}^N{x\over x^j-1}\;,\nonumber\ee 
\be 
G_g(z,x)=  { \cal Z}_{1,\infty}(-z/x,x) = 1+\sum_{N=1}^{\infty} z^N x^{gN(N-1)/2}\prod_{j=1}^N{1\over x^j-1}\;,
\nonumber\ee }
then $y_+ (1) =  { \cal Z}_{1,q}/ { \cal Z}_{2,q}$ in (\ref{coeffbis})  yields  the $q\to\infty$ identity}
%\iffalse 
%\be \sum_{n=1}^{\infty} z^n \sum_{l_1, l_2, \ldots, l_{j}\atop { \rm g-composition}\;{\rm of}\;n}\;c_g(l_1,l_2,\ldots,l_{j} )\;x^{\sum_{i=1}^j(i-1)l_i}=\ln \bigg({1+\sum_{N=1}^{\infty} z^N x^{gN(N-1)/2+N}\prod_{j=1}^N{1\over x^j-1}\over 1+\sum_{N=1}^{\infty} z^N x^{gN(N-1)/2}\hskip 0.4cm\prod_{j=1}^N{1\over x^j-1}}\bigg)\label{comfort}\ee\fi
\be 
\sum_{n=1}^{\infty} z^n \bb\sum_{l_1, l_2, \ldots, l_{j}\atop \text{g-composition}\;{\rm of}\;n} c_g(l_1,l_2,\ldots,l_{j} )\;x^{\sum_{i=1}^j(i-1)l_i}=\ln {{H_g(z,x)\over G_g(z,x)}} \;.
\label{comfort}\ee
When $g=1$, i.e., Fermi statistics and Pauli exclusion, the $c_1(l_1,l_2,\ldots,l_{j} )$ are non vanishing  for the sole composition $l_1=n$ with $c_1(n)=1/n$.  In this case (\ref{comfort}) trivially reduces to  $\sum_{n=1}^{\infty} {z^n/n}=-\ln(1-z)$.

When $g=2$, writing (\ref{nest}) for $z \to -z/x$ and $s(k)=x^k$ in the form
\be
{1\over y_+ (k)} = {1\over 1-{z x^{k-1} \over y_+ (k+1)}}
\nonumber\ee
and solving it iteratively for $y_+ (1)$ it yields as a solution the Ramanujan continuous fraction.
Exponentiating (\ref{comfort}), then, yields the identity 
 \iffalse \be
 \exp\bigg({ \sum_{n=1}^{\infty} z^n \sum_{l_1, l_2, \ldots, l_{j}\atop { \rm composition}\;{\rm of}\;n}\;c_2(l_1,l_2,\ldots,l_{j} )\;x^{\sum_{i=1}^j(i-1)l_i}}\bigg)={H_2(z,x)\over G_2(z,x)}=\frac{1}{\frac{-z}{\frac{-zx}{\frac{-zx^2}{\frac{-zx^3}{\frac{-zx^4} + + 1}+1}+1}+1}+1}
\nonumber\ee
\fi
\be
 \exp\bigg({ \sum_{n=1}^{\infty} z^n \sum_{l_1, l_2, \ldots, l_{j}\atop { \rm composition}\;{\rm of}\;n}\;c_2(l_1,l_2,\ldots,l_{j} )\;x^{\sum_{i=1}^j(i-1)l_i}}\bigg)={H_2(z,x)\over G_2(z,x)}=\frac{1}{1+\frac{-z}{1+\frac{-zx}{1+\frac{-zx^2}{1+\frac{-zx^3}{1+\frac{-zx^4}{1+\ldots}}}}}}
\nonumber\ee
{The LHS expresses the log of the Ramanujan fraction in terms of} $c_2(l_1,l_2,\ldots,l_{j} )$.
 
For a general $g$,  exponentiating  (\ref{comfort}) gives
\iffalse \be
 \exp\Bigg[{ \sum_{n=1}^{\infty} z^n \bb\bb\sum_{l_1, l_2, \ldots, l_{j}\atop \text{ g-composition}\;{\rm of}\;n}
 \bb c_g(l_1,l_2,\bb\ldots,l_{j} )\;x^{\sum_{i=1}^j(i-1)l_i}}\Bigg]={H_g(z,x)\over G_g(z,x)}
 \nonumber\ee
 {\color{green} [This is essentially restating (\ref{comfort}), decide if we keep]}
  giving\fi
an alternative expression of the ratio ${H_g(z,x)/ G_g(z,x)}$ in terms of the $c_g(l_1,l_2,\ldots,l_{j} )$ as a $g\ge 2$-generalisation of the OEIS coefficients  A227532 and A227543. Note, however, that in the case $g>2$ the iterative
solution of (\ref{atreat}) for $1/y_+$ leads to a complicated expression and does not yield any simple continuous
fraction as in the $g=2$ case.

\iffalse

so that (\ref{comfort}) rewrites as 
\be \sum_{n=1}^{\infty} z^n \sum_{l_1, l_2, \ldots, l_{j}\atop { \rm g-composition}\;{\rm of}\;n}\;c_g(l_1,l_2,\ldots,l_{j} )\;x^{\sum_{i=1}^j(i-1)l_i}=\ln {H_g(z,x)\over G_g(z,x)}\nonumber\ee
\noindent 

When $g=1$,  the ratio $H_1(-x,x)/G_1(-x,x)$  reduces to 
\be {H_1(-x,x)\over G_1(-x,x)} =\frac{1}{x+1}\nonumber\ee
 When $g=2$,  the ratio $H_2(-x,x)/G_2(-x,x)$  rewrites as Ramanujan continuous fraction
 \be {H_2(-x,x)\over G_2(-x,x)} =\frac{1}{\frac{x}{\frac{x^2}{\frac{x^3}{\frac{x^4}{\frac{x^5} + + 1}+1}+1}+1}+1}\nonumber\ee
 
  It would certainly be interesting to see if any matter  arises  along the lines just discussed for  cases beyond $g=2$.

\bea \sum_{k=1}^{\infty}n(k)&=&- \sum_{k=1}^{\infty}\sum_{n=1}^{\infty}\;(-1)^n \;(z x^{ k})^n\;\sum_{{l_1, l_2, \ldots, l_{j}\atop { \rm composition}\;{\rm of}\;n}} n\;c_2(l_1,l_2,\ldots,l_{j} ) \; x^{\sum_{i=1}^j(i-1)l_i}\nonumber\\ &=& -\sum_{n=1}^{\infty}\;(-1)^n \;z^n {x^n\over 1- x^n}\;\sum_{{l_1, l_2, \ldots, l_{j}\atop { \rm composition}\;{\rm of}\;n}}\; n\;c_2(l_1,l_2,\ldots,l_{j} ) \; x^{\sum_{i=1}^j(i-1)l_i}\nonumber\eea
\fi

\section{Conclusions}

{The analysis and results of this work could be applied to other systems, and in particular to the 
Calogero-Sutherland model of particles with an inverse sine
squared two-body potential, which can be viewed as exclusion-$g$ particles on the circle.
The spectral function for this model is (setting the length of the circle to $2\pi$)
\be
 s(k) = e^{-\beta k^2 /2} ~, ~~~ k= 0, \pm 1, \pm 2,\dots 
 \nonumber\ee
with $k$ interpreted as the discrete momentum of exclusion particles (also referred to as ``pseudomomentum'').
The spectrum is symmetric and unbounded in both ends, so forward and backward effective functions ${y_\pm} (k)$ are essentially
equivalent, satisfying $y_+ (k) = y_- (-k)$, a relation that survives the introduction of a symmetric cutoff $|k| \le q$.
We could also impose an asymmetric restriction to the spectrum, such as $k \ge 0$, which actually corresponds to a
symmetry-reduced Calogero-Sutherland model with inverse sine and inverse cosine squared potentials plus an additional
one-body interaction with the point at the origin.

The full Calogero-Sutherland spectrum can be treated by separating in the $N$-body partition function the (unrestricted)
sum over the smallest momentum $k_1$. The sum of the remaining momenta can be cast as the grand partition function of
$N-1$ particles with spectral function $s(k) = e^{-\beta(k_1 + g + k)^2}$ with $k \ge 0$. This makes the spectrum bounded
to the left, but creates the added complication of having to sum at the end over $k_1$. We defer any further
discussion of the microscopic statistics of the Calogero-Sutherland model to future work.
}

{In both the Calogero and LLL-anyon systems we can define ``quasihole'' excitations, corresponding to minimal ``gaps'' in a
completely filled state, that behave as particles with exclusion statistics $1/g$. We can define similar excitations in the microscopic (discrete) case, starting from a maximally filled many-body state and moving all particles to the right
of a marked particle by one level, or to the left by minus one level, therefore creating a quasihole. All the standard
features of quasiholes emerge in this picture: the removal of one particle creates $g$ quasiholes, identifying them as
$-1/g$ particle each, they ``move'' in the energy spectrum by steps of $g$, therefore corresponding to an $1/g$ dilution
of the density of states, and in a macroscopic span of $K$ successive levels a number $K$ of quasiholes 
can be``packed'' together. Since the effective
density of states has been decreased by a factor of $1/g$, this last property identifies quasiholes as particles with $1/g$ exclusion
statistics.

Several qualitatively new features arise, however, in the discrete setting: there are, now, $g$ distinct
``fully filled'' states, related to each other through a common shift of all particles by $1,2,\dots g-1$
levels, {so these states are not perturbatively connected}. Further, each quasihole's energy becomes a nontrivial combination of several energy levels. The exact properties
of quasiholes, and a potential reformulation of the statistical mechanics of the system in terms of them, remain
fascinating objects of further study.}

Exclusion statistics are related to the generating functions for the algebraic area counting of lattice walks \cite{latt}, a connection that arises from the
matrix determinant representation as exposed in section {\bf\ref{3}}. The Calogero model results, in particular, are
related to a set of directed walks termed Dyck or Lukasiewicz paths. This connection has been exploited to derive results
for such walks \cite{emeis} but it can also be used in reverse: the generating functions of generalized types of walks
can be considered as effective descriptions of particular exclusion-type statistics. The identification of such statistics,
or of the walks corresponding to other known types of statistics, is an interesting topic for investigation.

The most interesting possible application of our results is, however, in situations of physical significance, such as
fractional quantum Hall and related systems. This would require measurements that probe the microscopic quantities
studied in this work, such as, e.g., the single-level mean occupation numbers $n_k$, looking for
possible signatures within recent experimental results.  Also  LLL-anyon statistics  at maximal filling is encoded in  Laughlin type wavefuctions (\ref{laugh})  describing bulk particles with exclusion  $\alpha$, while experimentalists \cite{exp}  observe quasiholes with statistics $1/\alpha$.
Identifying possible  measurements allowing for an experimental confirmation of bulk particles with  exclusion $\alpha$
remains an important subject for further theoretical research and experimental developments.

\section{Appendix}

To derive (\ref{nss}), we use the recursion (\ref{recurk}) to rewrite ${\cal Z}_{k+1,q}={\cal{Z}}_{k+2,q}+z s(k+1){\cal{Z}}_{k+3,q}$ so that (\ref{neni}) becomes
  \be 1-n_k-n_{k+1}={{\cal Z}_{1,k-1}{\cal Z}_{k+2,q}\over  {\cal Z}_{1,q}}\;.\label{nenibis}\ee Then from (\ref{neni}) and (\ref{nenibis}) get
  \be {{\cal Z}_{k+2,q}\over{\cal Z}_{k+1,q}}={1-n_k-n_{k+1}\over 1-n_k}\label{coco}\ee
  Likewise from (\ref{nenibis}) get
  \be 1-n_{k-1}-n_{k}={{\cal Z}_{1,k-2}{\cal Z}_{k+1,q}\over  {\cal Z}_{1,q}}\ee
  or, again using (\ref{neni}),
  \be {1-n_{k-1}-n_{k}\over n_k}={{\cal Z}_{k+1,q}\over zs(k){\cal Z}_{k+2,q}}\;.\label{cocobis}\ee
  Multiplying (\ref{coco}) and (\ref{cocobis}) we obtain (\ref{nss}).

The proof of the corresponding relation for $g>2$ proceeds along similar lines, requiring now a ``telescoping'' product
relation to eliminate the various intermediate ${\cal Z}_{k,k'}$, and we leave it as an exercise to the reader.

\end{document}